\documentclass[a4paper,fleqn,usenatbib,useAMS]{mnras}

\usepackage{graphicx}	% Including figure files
\usepackage{amsmath}	% Advanced maths commands
\usepackage{amssymb}	% Extra maths symbols
\usepackage{multicol}        % Multi-column entries in tables
\usepackage{bm}		% Bold maths symbols, including upright Greek
\usepackage{pdflscape}	% Landscape pages
\usepackage{journals}

\def\HII{H~{\sc ii}\, }

\usepackage[T1]{fontenc}
\usepackage{ae,aecompl}

\usepackage{txfonts}
\title{Tidal bridge and tidal dwarf candidates in the  interacting system Arp194}
\author[A. Zasov et al.]{Anatoly V. Zasov$^{1,2}$\thanks{E-mail:
zasov@sai.msu.ru}, Anna S. Saburova$^1$, Oleg V. Egorov$^1$, Viktor L. Afanasiev$^3$ %\footnotemark[1]\thanks{This file has been amended to highlight the proper use of \LaTeXe\ code with the class file. These changes are for illustrative purposes and do not reflect the original paper by A. V. Raveendran.}
\\
$^1$ Sternberg Astronomical Institute, Moscow M.V. Lomonosov State University, Universitetskij pr., 13,  Moscow, 119991, Russia\\
$^2$ Faculty of Physics, Moscow M.V. Lomonosov State University, Leninskie gory 1,  Moscow, 119991, Russia \\
$^3$ Special Astrophysical Observatory, Russian Academy of Sciences, Nizhniy Arkhyz, Karachai-Cherkessian Republic 357147, Russia \\
}

\begin{document}
\label{firstpage}
\pagerange{\pageref{firstpage}--\pageref{lastpage}} \pubyear{2016}
\maketitle

\begin{abstract}

Arp194 is a system of recently collided galaxies, where the  southern galaxy (S) passed through the gaseous disc of the  northern galaxy (N) which in turn consists of two close components. This system is of  special interest due to the presence of regions of active star-formation in the bridge between galaxies, the brightest of which (the region A) has a size of at least 4 kpc.  We obtained three spectral slices of the system for different slit positions at the 6-m telescope of SAO RAS. We estimated the radial distribution of line-of-sight velocity and velocity dispersion as well as the intensities of emission lines and oxygen abundance $12+\log(\mathrm{O/H})$. The gas in the bridge is only partially mixed chemically and spatially: we observe the O/H gradient with the galactocentric distances both from S and N galaxies and a high dispersion of O/H in the outskirts of N-galaxy. Velocity dispersion of the emission-line gas is the lowest in  the star-forming sites of the bridge and exceeds 50-70 km/s in the disturbed region of N-galaxy. Based on the SDSS photometrical data and our kinematical profiles we measured the masses of stellar population and the dynamical masses of individual objects. We confirm that the region A is the gravitationally bound tidal dwarf with the age of $10^7 - 10^8 $ yr, which is falling onto the parent S- galaxy. There is no evidence of the significant amount of dark matter in this dwarf galaxy. 

\end{abstract}


\section{Introduction}
Arp 194 = VV126 is a tightly interacting system containing two main galaxies with active star-formation -- southern (S) and northern (N) ones, separated by $\sim 30$ kpc (in projection).  N-galaxy, in turn, consists of two apparently merging galaxies (Na-component possessing the distinct spiral arms and the less noticeable and redder component Nb, projected onto the eastern spiral arm of Na-galaxy).  The peripheral structure of the  galaxies, especially of the northern ones, is perturbed by the recent close approach or direct collision of N and S-galaxies. The best high-resolution images of this system may be found in the HST archive data.\footnote{The  color image may be found at the HST web-page: http://hubblesite.org/gallery/album/galaxy/interacting/pr2009018a}.

 Arp194 is of special interest due to the presence of the chain of star-formation regions between the galaxies which seldom occurs in interacting systems.  Indeed, if galaxies retained their integrality after the close passage, the extended regions containing young stars are usually observed either in long tidal tails or inside the galactic discs, but not in the bridge between galaxies. Among the Arp galaxies the extended islands of star-formation between the closely spaced systems are noticeable only in Arp269, Arp270 and, probably, in Arp59 where the interaction involves several galaxies. 

\begin{figure*} 
\includegraphics[width=11.5cm]{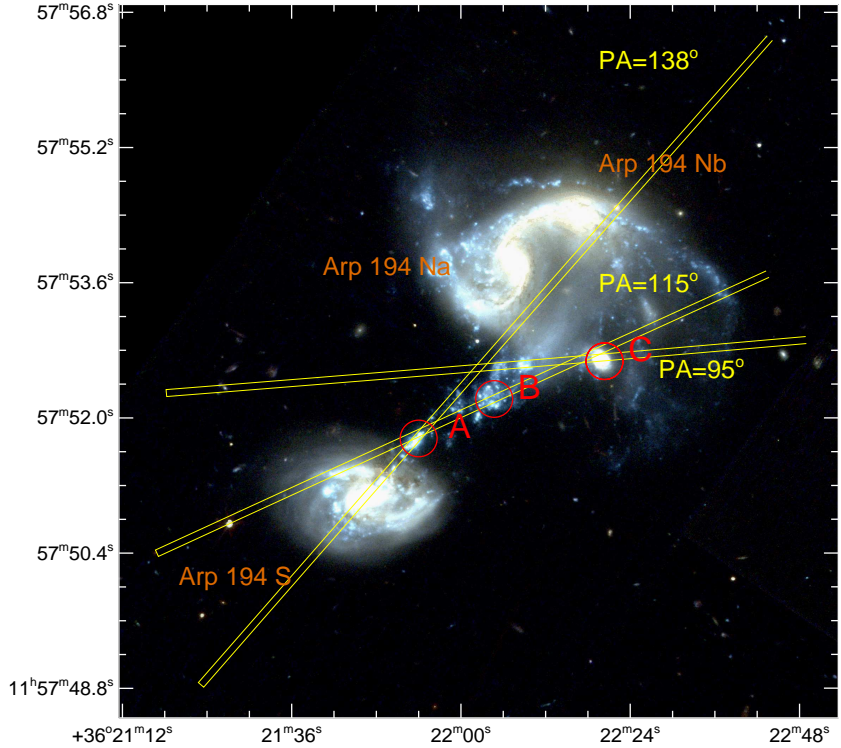}
\caption{The HST composite colour image in BVI bands  with overplotted  positions of the slits and regions of star-formation.}
\label{fig1}
\end{figure*}

The location of individual objects and their designation adopted in the current paper as well as the positions of the slit are shown in Fig. \ref{fig1} superimposed on the HST composite colour image. 

The mean system velocity of the components of Arp194 is close to 10500 $\textrm{km~s$^{-1}$}$, the difference between the central velocities of S- and N- galaxies does not exceed 30--40 $\textrm{km~s$^{-1}$}$. According to \cite{BushouseStanford1992} their stellar magnitudes in K band are 12.7 (S) and 12.1 (N), which correspond to the luminosities $3.6\cdot10^{10}(S)$ and $6.3\cdot10^{10}$  (N) in solar units adopting the distance 144 Mpc ($H_0=73\ \mathrm{ km\ s^{-1}\ Mpc^{-1}}$). Hence their NIR luminosities and consequently the stellar masses are comparable to that of the Galaxy.

The first spectral observations of the system were conducted at 6-m telescope BTA at the dawn of its operation by Karachentsev and Zasov in the end of 70-s using the UAGS spectrograph and the image intensifier  with photographic registration of spectra. The spectrograms were reduced by V. Metlov and published in \cite{Metlov1980}. The distribution of line-of-sight velocities of emission gas along the three spectral slices obtained in the cited paper unveiled the complexity of motions of gas and the presence of giant H{\sc ii} regions with absolute stellar magnitudes of -15 -- -17. A total dynamical mass of the system according to \cite{Metlov1980}  exceeds $10^{11}M\odot$.  A comparison of the velocity estimates with our measurements (this work) revealed a significant difference of some local values  (up to several tens of $\textrm{km~s$^{-1}$}$), mostly in the regions of strong emission. However, the overall manner of distribution of gas velocity along the slits from \cite{Metlov1980} agrees with the later estimates. 

More elaborate study of the system Arp194 was carried out by \cite{Marziani2003}  using the results of spectral and photometrical observations at the 2.1 m telescope. These authors obtained the spectral slices for two slit orientations and concluded that the unusual morphology of the system is related to the passage of the more compact southern galaxy through the disc of the northern one several hundred Myr ago, which caused the formation of the sequence of  bright emission islands between the galaxies and the apparently expanding ring structure around the northern galaxy (a collisional ring) which contains gas and star-forming regions. However, the detailed image of the system obtained later with HST demonstrates more complex picture: the arc of star-forming regions surrounding the system from the north side almost coincides with the spiral arm of Na-galaxy which is deformed and highly elongated to the west, at the opposite side there are no large-scale structures that could be attributed to the ring. \cite{Marziani2003} gave the convincing reasoning that the emission regions between the galaxies represent the `blobs of stripped gas due to the interpenetrating encounter' of southern and northern  components of the system. In other words, they resulted from the star-burst in the cold gas lost by galaxies during the collision. The absence of noticeable radiation in J and K-bands speaks in favour of young age of these emission bright regions. \cite{Marziani2003} estimated the age of the brightest stellar island A which is close to S-galaxy (see Fig.\ref{fig1}): $T=7\cdot10^7$ yr and concluded that it falls into the central part of the galaxy, which is in a good agreement with the conclusions of the current paper.

The goal of current paper is the study of the individual star-forming regions on the periphery of the galaxies and between them, which could be considered as candidates to the tidal dwarfs parallel with the analysis of dynamics and chemical composition of gas in the system. 

The paper is organized as follows: in Sec.~\ref{Obs} we present the results of our long-slit spectral observations of Arp194 conducted at BTA, including the description of data reduction and the kinematic and chemical abundance radial profiles. Sections \ref{Discussion} and \ref{conclusion} are devoted to the analysis of the obtained data and present general results. In these sections we also involve the photometric data from SDSS.

\section{Observations}\label{Obs}
\subsection{Data reduction}
We performed the long-slit spectral observations of Arp194 in 2013-2016 using the spectrograph SCORPIO-2 \citep{AfanasievMoiseev2011} mounted at the
prime focus of the 6-m Russian telescope BTA at Special
Astrophysical Observatory of the Russian Academy of Sciences (SAO RAS). Three different positions of the slit were used, so that each one crossed a certain compact object of the system. The slit with positional angle PA=115\degr ~is close to that used by  \cite{Marziani2003} with PA=118\degr, ~but in our  case the slit width is more than twice narrower (about 1'').  The log of the observations is given in Table \ref{log}. The positions of the slits are shown in Fig. \ref{fig1}. We used the grism VPHG1200@540 which covers the spectral range 3600-7070 \AA ~and has a dispersion of 0.87 \AA ~pixel$^{-1}$. 
\begin{table}
\caption{Log of observations}\label{log}
\begin{center}
\begin{tabular}{cccc}
\hline\hline
Slit PA & Date & Exposure time& Seeing \\
   ($^o$)  &  &     (s) &        (arcsec) \\
\hline
115&10.02.2013&3600&1.5 \\
138&24.03.2015&1800&4.2 \\
95&09.03.2016&900& 1.2 \\
\hline\hline
\end{tabular}
\end{center}
\end{table}

The data reduction was performed using the \textsc{IDL}-based reduction pipeline. Firstly, we subtracted the bias, performed the flat-fielding and removed the hits of cosmic rays. The wavelength calibration was made using the spectrum of He-Ne-Ar lamp. The uncertainties of the wavelength solution are within 0.1 \AA ~for all spectra. After the wavelength calibration we performed the linearization and summation of the spectra obtained for the same slit position. For measuring the night sky spectrum we used the peripheral regions of the slits where the galaxy light contribution is negligible. Then we transformed the peripheral night sky spectrum into the Fourier space and  extrapolated it onto the galaxy position by using the polynomial representation at a given wavelength. 

\begin{figure}                                                               
 \centering                                                                   
 \includegraphics[width=\linewidth]{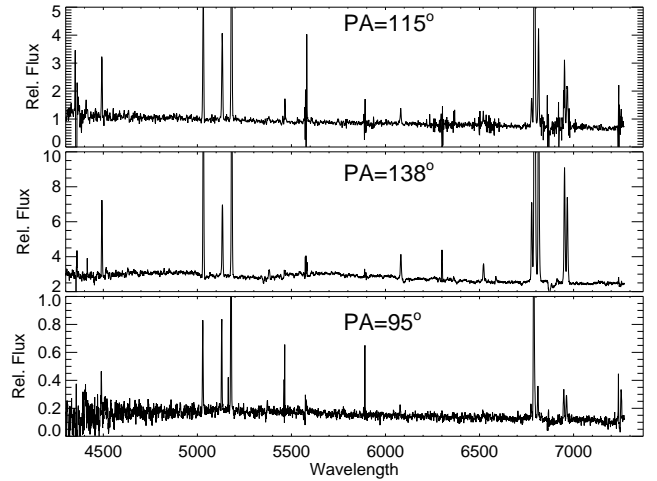}       
  \caption{ The sky-subtracted spectra for three slit orientations, integrated along the slits. The flux is in units of $10^{-15} erg/cm^2/sec/$\AA, the wavelength is in \AA, (non-corrected for the redshift).  The most notable lines are $H\gamma$ 4340, $H\beta$ 4861,  [OIII] 4959,5007 and $H\alpha$	6563.}
\label{spectra} 
\end{figure}

After the sky subtraction we performed the flux calibration using the spectro-photometrical stellar standards: HZ44, G191B2B, Feige56 and BD+33d2642. The sky-subtracted spectra integrated along the regions crossed by slit for three positions of the slit are shown in Fig. \ref{spectra}.

In order to estimate the parameters of the instrumental profile of spectrograph we analyzed the spectra of twilight sky, taken in the same night or in the adjacent nights. These spectra were divided into several bands where the signal was summarized in order to increase the signal-to-noise ratio. After that we fitted the spectrum in each band by a broadened high-resolution spectrum of the Sun. The resulting parameters of instrumental profile and their variation both with the wavelength and along the slit were utilized in the spectral fitting by convolving the profile with the grid of stellar population models.

To extract the emission lines from the stellar background  and to obtain the properties of stellar population we firstly binned the spectra using the adaptive binning algorithm in order to achieve minimal signal-to-noise ratio $S/N=10$. After that we used the NBursts full spectral fitting technique \citep{Chilingarian2007}. This technique enables one to fit the observed spectrum in the pixel space against the population model convolved with a parametric line-of-sight velocity distribution. We utilized the high resolution stellar PEGASE.HR \citep{ LeBorgne2004}
 simple stellar population (SSP) models based on the ELODIE3.1 empirical stellar library. NBursts technique allowed us to obtain the parameters of stellar population (velocity, velocity dispersion, age and metallicity)  by means of non-linear minimization of the quadratic difference chi-square between the observed and model spectra. After that we subtracted the models of stellar spectra from the observed ones and got the pure emission spectra. We fitted the emission lines by Gaussian distribution and derived the velocity and velocity dispersion of the ionized gas. 
 
 Unfortunately, the signal-to-noise ratio in our spectra is not sufficient to measure fluxes of the sensitive to electron temperature faint emission lines [OIII] 4363 \AA\, and [NII] 5755 \AA. Thus we were unable to obtain the chemical abundances with `direct' $T_e$ method. Instead we used the strong-lines empirical calibrations to estimate the gas oxygen abundance.
 
 Due to the well-known and still unresolved problem of discrepancy between the abundance estimates based on the  different empirical calibrations \citep[see, e.g., ][]{KewleyEllison08, Lopez-Sanchez12} we decided to use several methods to compute the oxygen abundance. One of them -- izi \citep{Blanc15} -- is calibrated with theoretical photoionization models and hence it might provide the elevated results in comparison with the `direct' $T_e$ method. This method derives the oxygen abundance through Bayesian inference using the fluxes of all measured emission lines.  Two another methods we used are calibrated with the sample of  \HII regions with reliable measurements of electron temperature. One of them is C-method \citep{Pilyugin12}, which derives the oxygen abundance by comparison of combinations of strong-line intensities measured in the spectrum with the sample of well-studied \HII regions with known metallicity. The second method -- O3N2 \citep{O3N2}, is based on the ratio of the strong emission lines [OIII]5007 and [NII]6584: $\mathrm{O3N2 = ([OIII]/H\beta)/([NII]/H\alpha)}$.
 
The oxygen abundances derived with three  methods mentioned above show similar behaviour, however the values obtained with izi method are typically higher with respect to the other methods. In our analysis we prefer to focus on the O3N2 approach, but in the following figures we present the radial profiles of oxygen abundance derived with all three methods.
 
\subsection{Kinematical profiles}\label{kin_prof}
\begin{figure} 
	\includegraphics[width=\linewidth]{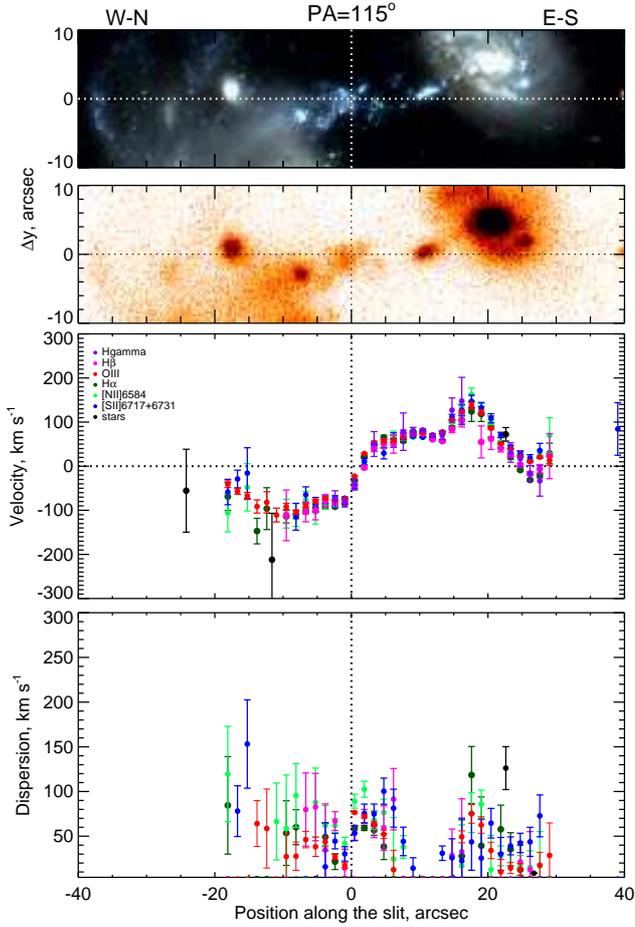}
	\caption{The radial kinematical profiles obtained for PA=115\degr. Two panels on top correspond to reference image from HST and 
		SCOPRIO-2. Next two panels show the velocities and
		velocity dispersions. The zero point for the velocity $V_{sys}=10440$ $\textrm{km~s$^{-1}$}$ which is adopted for all profiles. Black symbols correspond to stellar kinematics, colour (grey in printed version) circles -- to
		different emission lines (their designations are given in figure).}
	\label{fig2}
\end{figure}

\begin{figure} 
	\includegraphics[width=\linewidth]{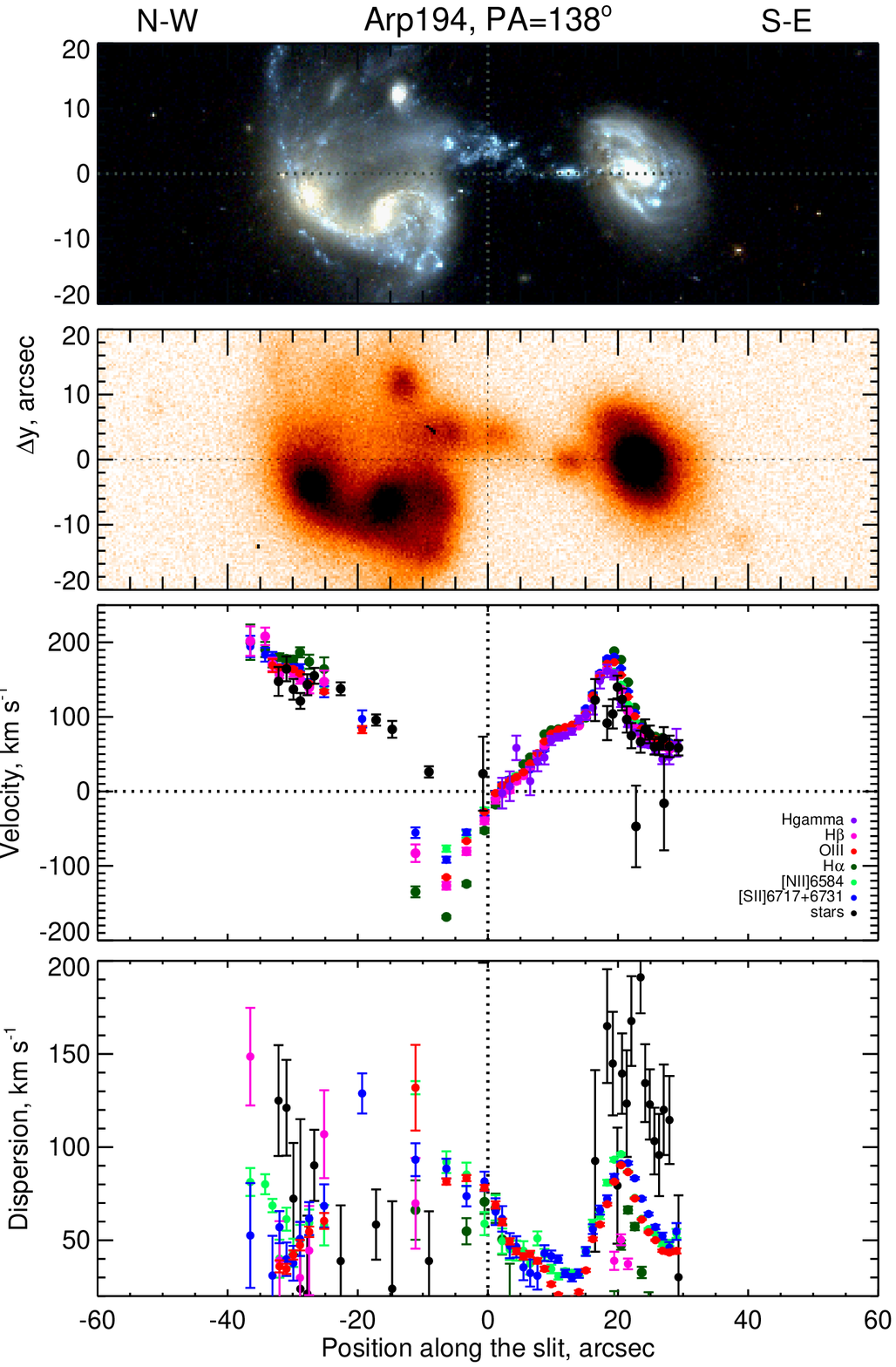}
	\caption{The same as in Fig. \ref{fig2}  for PA=138\degr. 
	}
	\label{fig3}
\end{figure}

\begin{figure} 
	\includegraphics[width=\linewidth]{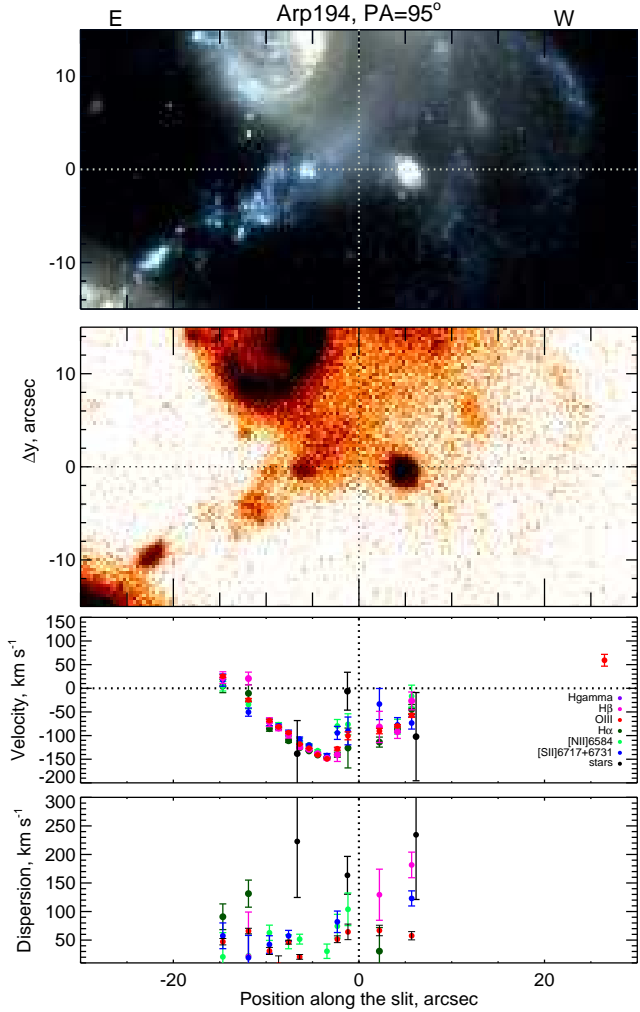}
	\caption{The same as in Fig. \ref{fig2} for PA=95\degr. 
	}
	\label{fig4}
\end{figure}

Figs. \ref{fig2}-\ref{fig4} show  the radial distribution of the line-of-sight velocity and velocity dispersion of ionized gas (coloured (grey in printed version) circles) and stars (black circles) for three positional angles. Two top panels correspond to the reference image from HST (above) and from BTA SCOPRIO-2 device (below). The horizontal line on each line-of-sight velocity profile corresponds to the mean velocity of the galaxies taken as  $V_{sys}=10440$ $\textrm{km~s$^{-1}$}$, zero point of horizontal axis is taken arbitrary.

We compared our estimates of the line-of-sight velocity along PA=115\degr ~with that from \cite{Marziani2003} and PA=118\degr ~and found a good agreement: the deviations do not exceed 15 $\textrm{km~s$^{-1}$}$ or remain within the error-bars when they exceed this value.

From Figs. \ref{fig2}, \ref{fig3} (on the right-hand side) one can see a distinct velocity gradient through S-galaxy evidently caused by its rotation. When the slit PA=138\degr passes through the visible edge of S-galaxy and comes to the region A,  velocity decreases abruptly by 90 $\textrm{km~s$^{-1}$}$ within the  interval of only 7 arcsec ($\sim 5$ kpc). Actually this change should be even more steep since the observed gradient is smoothed by the overlapping of S-galaxy and the region A, as it  was noted earlier by \cite{Marziani2003}. Within the region A the velocity changes less than by 40~$\textrm{km~s$^{-1}$}$. However, beyond this region the velocity shows a fast decrease in the direction of N-galaxy when the slit crosses the low surface brightness area containing a handful of emission knots. We denote this area  as the region B, it will be discussed below. 
%It should be noted that the change of the line-of-sight velocity is almost the same within this object for two positions of the slit (PA=115\degr ~and PA=138\degr). 
Finally, at the distance of 30-35 arcsec from the centre of S-galaxy the velocity gradient sharply changes its sign, as it is seen both at PA=138\degr and PA=115\degr. Then  the line-of-sight velocity begins to rise along the outskirts of the companions of N- galaxy (PA=138\degr) or remains nearly constant (PA=138\degr).  It seems that we have two systems of gas between galaxies, one of them is genetically related to N- galaxy, and the other one -- to its intruder. The area where the gradient changes  can be arbitrary attributed to the boundary between the gas systems of N- and S-galaxies.  Region A and probably region B  (see Fig. \ref{fig1}) are on the side of S- galaxy, the latter is situated near the conventional borderline. 

Inside of N-galaxy and in  its northern  surrounding we observe a  superposition of two motions of gas: the rotation responsible for the velocity gradient along the slit, and the non-circular motions of gas and stars which are naturally expected in a gravitationally perturbed system. None of our slit positions crosses the main body of the galaxy Na or Nb, so we have no measurements of their velocity dispersion. However in the case of Nb, the slit  PA=138\degr ~passed at the distance of only several kpc from its centre. The velocity dispersion of stars appears to be very high there: 100-130 $\textrm{km~s$^{-1}$}$ (it is higher only for stars in the central part of S-galaxy). Note that the line-of-sight velocity of stars changes along the slit in the region of N-galaxies not so steep as the velocity of gas. Apparently the stellar velocities better reflect the rotation of the disc of Na-galaxy than that of the gas. Indeed, while the velocities of stars and gas practically coincide in the inner part of Nb- galaxy  (left-hand side in Fig. \ref{fig3}), in the outer part of N-system the line-of-sight velocity of gas is by $\sim 100$ $\textrm{km~s$^{-1}$}$ lower than that of stars. Presumably the gas is not in the plane of stellar disc  in this area.

For PA=138\degr ~the stellar velocity dispersion appears to be the highest for the centre of S-galaxy, where it rises to $\sim 170$ $\textrm{km~s$^{-1}$}$. This value is quite typical for bulges of galaxies. The velocity dispersion of ionized gas in this region is roughly two times lower, which is quite expectable for the gas concentrated in the rotating disc. The lowest value of velocity dispersion of gas (approximately 30 $\textrm{km~s$^{-1}$}$) is observed for stellar island A.  Such value  commonly occurs in star-forming regions of discy galaxies. The velocity dispersion of gas increases beyond this island toward the boundary between the systems of N- and S- galaxies (X=-10 in Fig. \ref{fig3}), which can indicate the presence of gaseous streams initiated by the collision of gas subsystems.

\begin{figure} 
	\includegraphics[width=\linewidth]{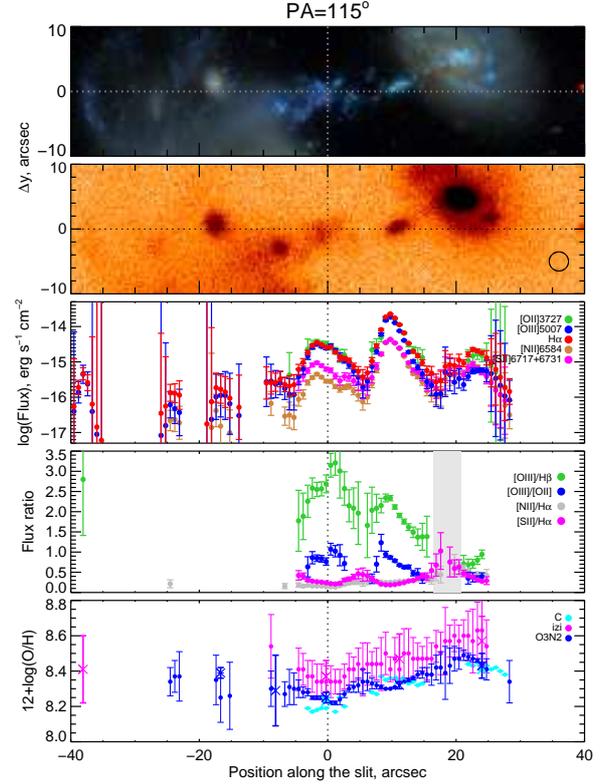}
	\caption{The distribution of oxygen abundance, emission lines fluxes in the log scale and flux ratios along the slit with PA=115\degr. Two panels on top are the  reference images taken from HST and obtained at
		SCOPRIO-2. Crosses at the bottom panel denote the mean values of oxygen abundance integrated over the extended regions. Grey shaded stripes highlight the regions with probably significant contribution of shock excitation, as it follows from the diagnostic diagrams.}
	\label{abund115}
\end{figure}

The radial profiles of velocity and velocity dispersion along the slit with PA=115\degr ~(see Fig. \ref{fig2}) demonstrate similar  behaviour as for PA=138\degr. One can see there a distinct velocity gradient when the slit crosses S-galaxy and after that a sharp decrease of the velocity and almost a plateau for region A,  followed by the decreasing of velocity towards  the boundary between the regions controlled by S- and N-galaxies. The minimal value of line-of-sight velocity is observed in the vicinity of region B, after that it experiences a little change and slightly grows near the C-region along the eastern periphery of northern galaxy. The measurements of velocity dispersion of stars and gas are more uncertain for this cut. In most regions the velocity dispersion of gas does not exceed 50 $\textrm{km~s$^{-1}$}$. A well defined minimum is observed  near the region B (X=0-- -5 in Fig. \ref{fig3}), after that the  dispersion grows, reaching  100-150 $\textrm{km~s$^{-1}$}$ near the region C (X = -15-- -20).

The third cut (PA=95\degr, Fig. \ref{fig4}) crosses small emission  regions between the galaxies in the  vicinity of Na and passes through the outer regions of compact object C on the west side of the system. The spectrum of this object contains, apart from the lines expected for the system Arp194, the strongly shifted lines $H_\alpha$, [NII], $H_\beta$ and [OII]. Their red-shift corresponds to the velocity of about 30000 $\textrm{km~s$^{-1}$}$. It gives evidences that the object C is a distant background galaxy (at the  HST image one can see its fuzzy spiral arms twisted clockwise). The line-of-sight velocity changes along the slit PA=95\degr ~by 150 $\textrm{km~s$^{-1}$}$ in the interval of 20 arcsec when the slit crosses the peripheral parts of N-galaxies and reaches its minimal value in the vicinity of the object C. The velocity dispersion of gas and stars in this area is surprisingly high ($\sim 150$ $\textrm{km~s$^{-1}$}$) and gas and stars have similar negative velocities there. It evidences in favour of the stream of matter towards the observer relatively to the mean velocity of the system. Note that for five out of six local estimates along the slit the measured stellar velocity dispersion is unusually high 
(100-200  $\textrm{km~s$^{-1}$}$), which clearly shows that here we observe stars lost by the disc of their parent galaxy.

\subsection{Oxygen abundance of ionized gas}

\begin{figure} 
	\includegraphics[width=\linewidth]{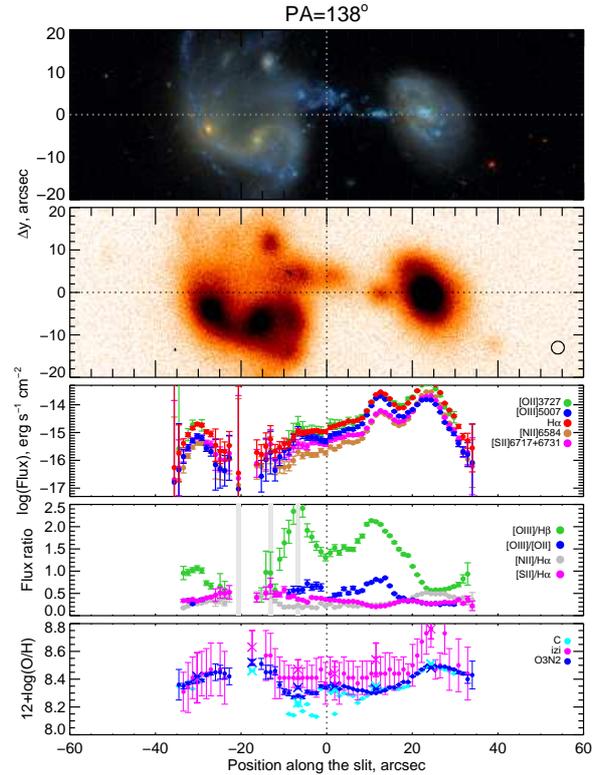}
	\caption{The same as in Fig. \ref{abund115}, but for PA=138\degr. }
	\label{abund138}
\end{figure}

\begin{figure} 
	\includegraphics[width=\linewidth]{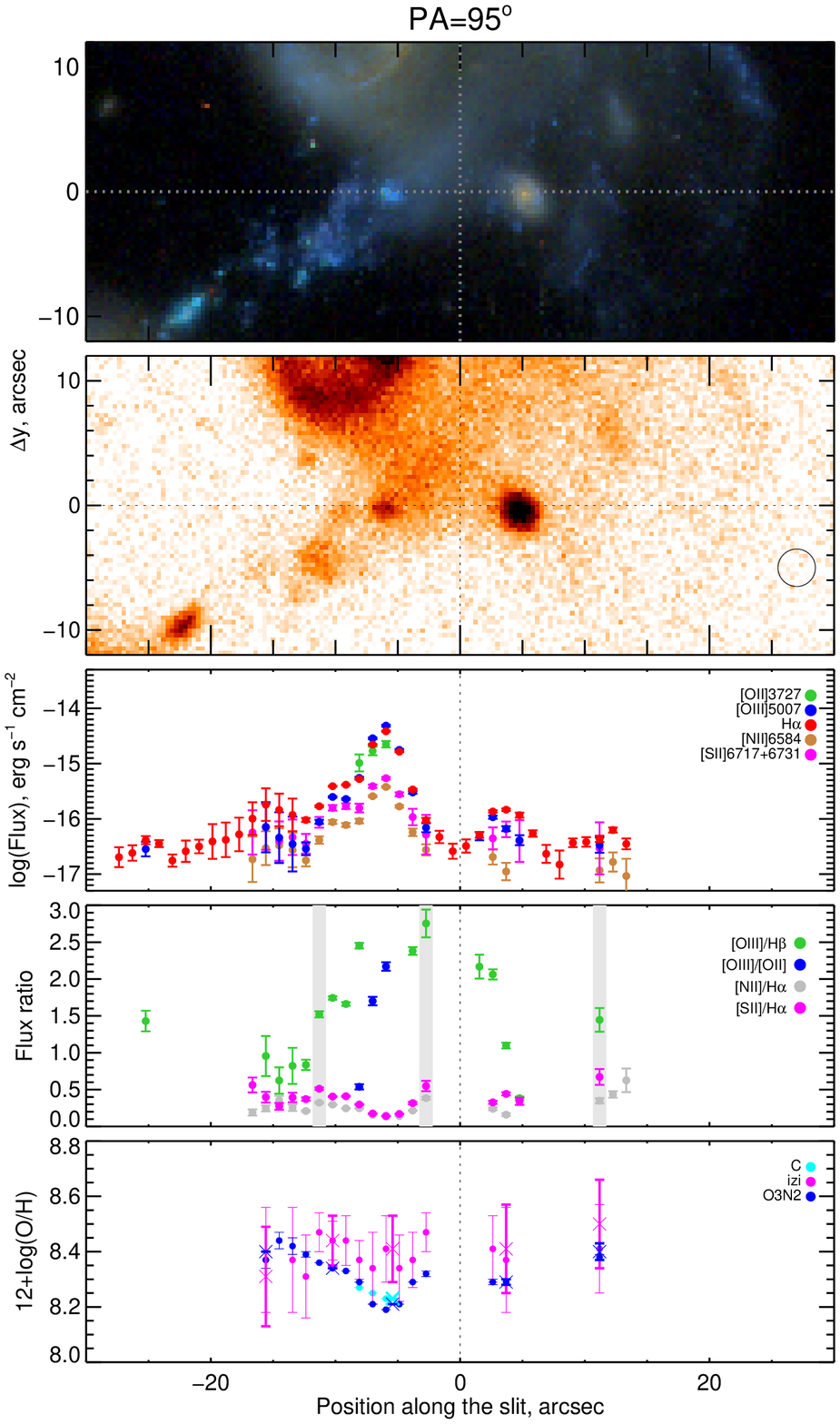}
	\caption{The same as in Fig. \ref{abund115}, but for PA=95\degr. }
	\label{abund95}
\end{figure}

\begin{figure} 
	\includegraphics[width=\linewidth]{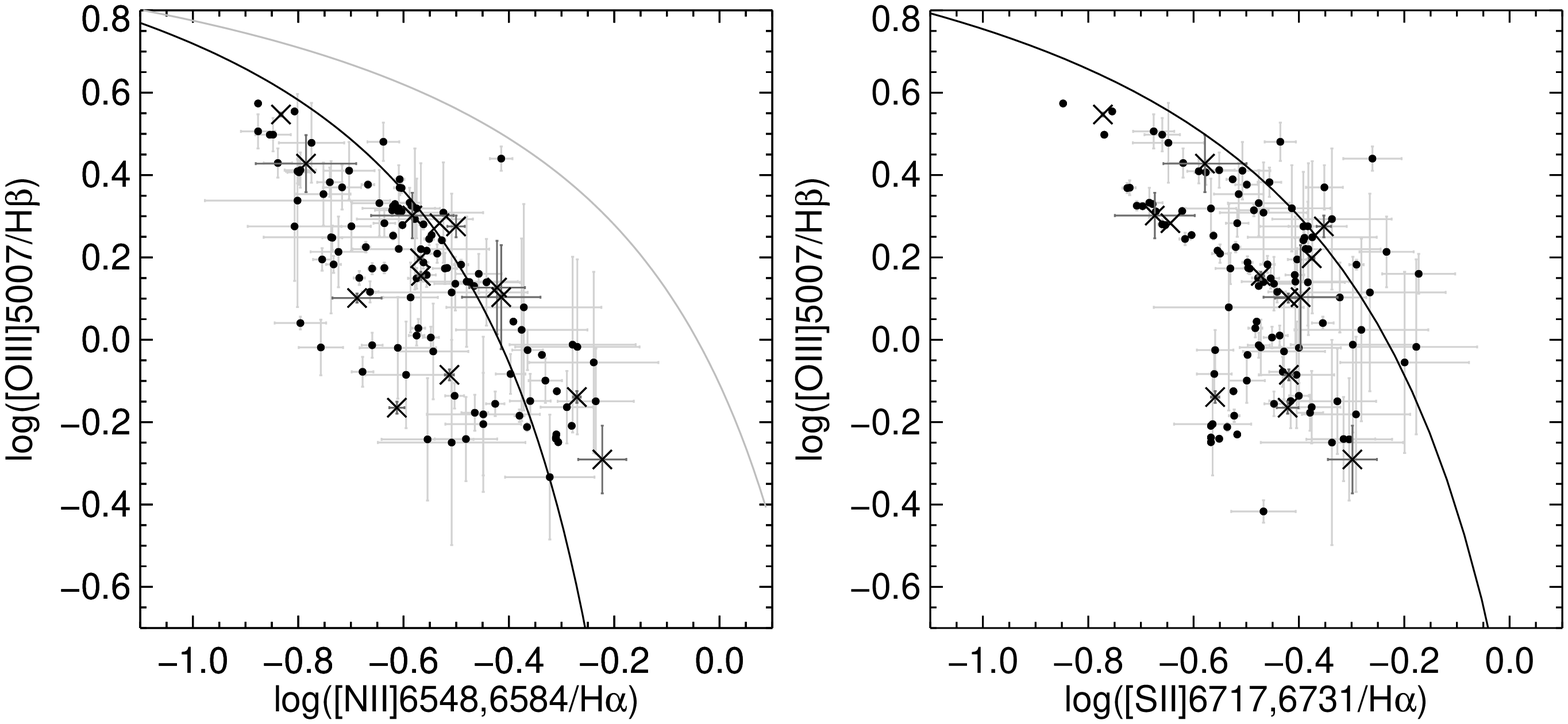}
	\caption{Diagnostic `BPT' diagrams $\mathrm{\log([OIII]/H\beta)}$ vs $\mathrm{\log([NII]/H\alpha)}$ (left) and $\mathrm{\log([SII]/H\alpha)}$ (right)   constructed for all three slit positions. Solid lines separate regions of pure photoionization excitation (below the black line), shock excitation (above both lines) and combined contribution of both mechanisms (between grey and black lines in the left-hand panel) (see \citet{Kauffmann03} and \citet{Kewley01}).}
	\label{BPT}
\end{figure}

The profiles of emission lines fluxes along the slits, their ratios and oxygen abundance $12+\log(\mathrm{O/H})$ are shown in Figs.~\ref{abund115}--\ref{abund95}. In order to distinguish between the pure photoionized regions and those with significant contribution of shock excitation we plotted the computed flux ratios on the classical diagnostic [OIII]/H$\beta$ vs [NII]/H$\alpha$ and [SII]/H$\alpha$ `BPT' diagrams \citep*{BPT} shown in Fig.~\ref{BPT}.  The separation curves according to the models of \citet{Kewley01, Kauffmann03} are superimposed. Grey vertical stripes in the radial profiles of emission line fluxes in Figs.~\ref{abund115}--\ref{abund95} mark the regions where the diagnostic ratios indicate the significant portion of radiation from the shock excited gas. As it follows from these diagrams, in a majority of areas crossed by the slits the lines have a recombination nature of emission, and consequently are related to the massive stars. The shock excitation plays the most important role in the inner region of S-galaxy where the intensive star-formation occurs, initiated by non-circular gas motions, and in the regions of low surface brightness to the west from the pair of N-galaxies where there are no noticeable extended  star-forming regions and the emission is apparently related to the diffuse ionized gas.  This gas with low intensity of emission lines   is also characterized by the enhanced velocity  dispersion (see Sect. \ref{kin_prof}). It agrees with the conclusion made by \cite{Moiseev2012} for dwarf galaxies (and evidently having a more general nature) that most of the areas in galaxies with the highest velocity dispersion
belong to the diffuse low brightness gas.  
Note that in normal spiral galaxies the collisional excitation of  gas  may usually be found in  circumnuclear regions evidencing the AGN activity or in the diffuse gas above the galactic plane, that could be ionized by shock waves from SNe. One may propose that in the case of colliding galaxies а strong non-circular or slowly decaying high velocity turbulent motions may be responsible for the locally enlarged input of collisional excitation into the gas emission. The increased velocity dispersion in the regions with the enhanced role of collision excitation in our data favours this suggestion.

\begin{figure} 
	\includegraphics[width=\linewidth]{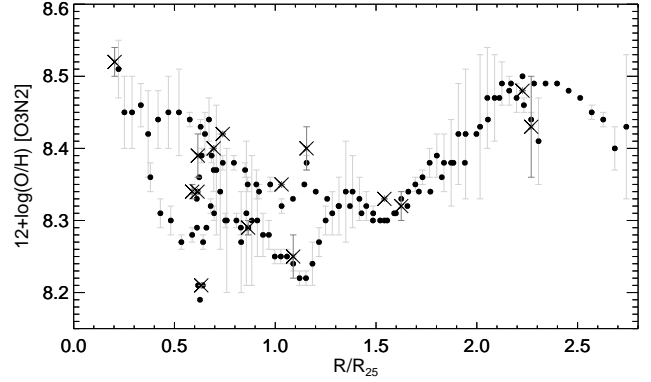}
	\caption{Variation of oxygen abundance estimated by O3N2 method with the distance from Na-galaxy along all three slits. As a zero-point at the x-axis we took the centre of Na galaxy, the radial scale is normalized to the photometric radius of Na galaxy $R_{25} = 19$~kpc. The centre of S-galaxy is located at $R/R_{25}\approx 2.3$.}
	\label{abund_grad}
\end{figure}

%\begin{figure} 
%	\rule{5cm}{3cm}
%	\caption{Oxygen abundance vs stellar mass dependence for galaxies in CALIFA survey (taken from \citealt{Califa2016}). Positions of the regions A, B, C are shown.}
%	\label{abund_CALIFA}
%\end{figure}

The chemical inhomogeneity of the system is evident from the presented profiles of oxygen abundance. Although the range of abundances within the system as whole is not too large (about 0.4 dex, or the factor 2.5), its distribution reveals some regularity. It contrasts with the case of post-collision system Arp 270,  also consisting of two recently collided star-forming galaxies, where the gas is chemically  well mixed all over the system  (\citealt{Zasovetal2015}). 

Regions between the galaxies possess the lowest oxygen abundance in comparison with the abundances of the main galaxies -- irrespective of the brightness of emission lines (see Figs.~\ref{abund115}--\ref{abund95}) This  is also clearly illustrated by the diagram where the oxygen abundances obtained for all three spectral cuts plotted against the distance from the centre of galaxy Na, arbitrary chosen as the origin of the coordinates (see Fig.~\ref{abund_grad}). The distance is normalized by the optical radius $R_{25}$ of Na-galaxy. A position of S-galaxy is about $R/R_{25}=2.3$.  The metallicity of gas in the main bodies of S and N- galaxies, at least in the areas crossed by the slit, is $\sim8.3 - 8.5$ (for O3N2 method) which is slightly lower than the solar value usually considered as  $ 12+\log(\mathrm{O/H})_\odot = 8.69$ \citep{Asplund2009}. However the galaxies remain within the range of metallicities expected for given luminosities (see e.g. \citealt{Califa2016}).

\section{Discussion}\label{Discussion}

\subsection{N- and S- galaxies}
As we noted in the Introduction, the northern galaxy consists of two close components -- Na and Nb, which are apparently connected by stellar bridge or by distorted spiral arm.  The component Nb is barely distinguishable in the visible light against the background of the deformed spiral arm of its companion, looking as the part of its spiral structure. Nevertheless Nb appears to be brighter than Na in the near infra-red (K-band). It indirectly indicates that the stellar mass of Nb is higher than that of Na. \cite{Marziani2003} described this object as the reddest one in the system: its colour index $B-R=1.17$. The elongated blue arc being apparently a spiral arm of Na, passes (in projection) through Nb without distortion. It could indicate that the Nb galaxy lays beyond the gas-dust disc of Na being observed through its outer parts. The extended spiral arm of Na galaxy is bordered by the uneven dust lane running along its inner edge. It gives evidence of the wave nature of spiral arm and allows to conclude that the angular velocity of the density wave is lower than that of the disc. The red colour of Nb agrees with the absence of the observed manifestation of young stellar population in this galaxy although partly it may be caused by the light scattering by dust in the disc of Na-galaxy. However the extinction correction measured by the ratio of intensities in H$\alpha$  and H$\beta$ is not high ($\sim 0.15$ mag).

The difference between the line-of-sight velocities of Na and Nb-galaxies is approximately 150 $\textrm{km~s$^{-1}$}$ (the velocity is higher for Nb). The presence of stellar `bridge' between  these galaxies shows that they undergo a stage of highly-developed merging. Thus the deformed shape of the galaxies and non-circular motions of gas between them is probably due to their interaction with each other rather than with S-galaxy. 

The measured velocities of emission gas allow to estimate roughly the dynamical masses\footnote{$M_{dyn}\approx V^2R/G$, where R is the radius of object, $V$ is its rotation velocity. For A and B regions this estimate can be considered only as an upper limit of mass, since this relation is valid only for gravitationally bound systems which in general is not evident.} of the system members and to compare their virial masses with the masses of their stellar population. We calculated the latter ones from the luminosity in $g$-band using the mass-to-light ratio $ M/L_g$ as the function of $g-r$ colour following the model relations of \cite{Belletal2003}. For the photometric analysis we used the images taken from SDSS survey. The resulting photometric and dynamical  estimates are given in Table \ref{tab_phot}.

Rotation of N-galaxy most clearly manifests itself in the velocity profile along PA=118\degr ~obtained by \cite{Marziani2003}. This cut passes close both to the centre of the galaxy and to its major axis. South-east side of the galaxy is approaching, and north-west side is receding with respect to its centre.  In the expected case of counter-clockwise rotation of the galaxy, its south-west part is the closest to the observer. The spectral slice demonstrates a distinct and practically linear variation of the line-of-sight velocity of gas $\Delta V$ amounting $\sim 200$ $\mathrm{km\ s^{-1}}$ within the radial distance $R=25$~arcsec (17 kpc) (see fig. 7 in  \citealt{Marziani2003}). Our spectral slice PA=138\degr ~passes nearly parallel to the major axis at some distance from the centre and demonstrates similar velocity profile.	Assuming the axis ratio of the disc is close to 0.5, as it follows from the image of Na-galaxy, one can give the rough estimate of its total mass: $M \approx ((\Delta V/2 \sin(\arccos(b/a)) )^2\cdot R/2G \approx 3\cdot10^{10} M_\odot$. This value is in a good agreement with photometric estimate of stellar mass of the two galaxies: $M_{ph}= M_a+M_b= (L_g\cdot(M/L)_g)_{Na} +(L_g\cdot(M/L)_g)_{Nb} = 2.9\cdot10^{10}M_\odot$ which gives evidences in favour of rotation of double galaxy Na+Nb as whole and excludes the mass domination of dark matter halo within the considered radial distance range. 

The mass of S-galaxy inside of its optical radius also can be found from the available spectral data. One of our slits  (PA=138\degr) intersects the centre of the galaxy, while the other one (PA=115\degr) cuts its peripheral regions. Our data for PA=138\degr ~agree satisfactorily with that of \cite{Marziani2003} for the slit crossing the centre with PA=145\degr ~up to $\sim 7$ kpc in projected plane. Farther from the centre their measurements have high error-bars. To estimate roughly  the mass of this galaxy one can suppose that the velocity of gas in the galaxy is close to circular one. This is not the case at least for the part of the disc faced to the region A where asymmetric  velocity  profile indicates the presence of non-circular motions, so we exclude this region for mass estimation.

In principle, velocities of circular rotation can be extracted from any spectral slice of a disc, when its  spatial orientation is known. Indeed, the observed line-of-sight velocity and radial coordinate $R$ of a given point is unambiguously determined by the circular velocity $V(R)$ and angle $\phi$ between the radius-vector of a given point and the major axis of a galaxy: 
\begin{equation}
\label{formula_v} V(R) = \frac{V_r(R)\sqrt{(sec^2(i)-\tan^2(i)\cos^2(\phi))}}{\sin(i) \cos(\phi)}
\end{equation}

\begin{equation}
\label{formula_r} R = R_\phi \sqrt{(\sec^2(i) - \tan^2(i) \cos^2 (\alpha))}
\end{equation}
Here  $ R_\phi $ is the radius in the sky plane, $V_r$ is the line-of-sight velocity corrected for the systemic velocity, $i$ is inclination of the disc. The analysis of r-band SDSS image showed that the shape of the outer isophotes of S-galaxy is best described by the inclination and position angles $i=58\degr$ and PA=$-9$\degr ~within the errors of a few degrees.

\begin{figure} 
\includegraphics[width=\linewidth]{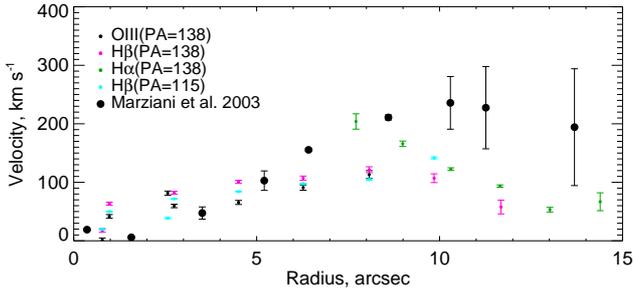}
\caption{The ionized gas  rotation curve of S-galaxy calculated for two different spectral slices (PA=115 \degr ~and PA=138 \degr) (small circles) with the overplotted Marziani et al. (2003) data (PA=145 \degr) (big black circles).}
\label{rc}
\end{figure}

In Fig. \ref{rc} we give the resulting estimates of rotation velocity of S-galaxy as a function of $R$. In spite of a large spread of values, evidencing the non-circular motions even in the southern half of a galaxy, the maximal velocity $V_m\approx 160-200\ \mathrm{km\ s^{-1}}$ is determined rather reliably within the radius we consider (11 arcsec $\approx$ 7 kpc). Thus the mass enclosed within the visible radius of the galaxy  is about $4.6\cdot 10^{10} M_\odot$ which agrees with the estimate of stellar mass of this galaxy (see Table \ref{tab_phot}) and the moderate relative mass of the dark halo ($M_{halo} \approx 1.4 M_*$).
\begin{table*}
\caption{The photometrical properties of the regions (1) -- Name; (2) -- Radius of the aperture; (3) --  $g$-band luminosity, the adopted distance is 144 Mpc for all regions except C, for which we used 400 Mpc; (4) -- the $(g-r)$ colour uncorrected for internal extinction; (5) -- the $(u-g)$ colour uncorrected for internal extinction; (6) -- the correction for internal extinction according to flux ratio H$\alpha$/H$\beta$; (7) -- Dynamical (virial) mass for the regions with measured velocity gradient; (8) -- the photometric stellar mass following from $(g-r)$ colour and the  model relation from \citet{Belletal2003} \label{tab_phot}}
\begin{center}
\begin{tabular}{cccccccc}
\hline\hline
Region & Radius & $L_g$& $(g-r)$& $(u-g)$ & $E(B-V)$&$(M)_\mathrm{dyn}$&$(M)_\mathrm{phot}$\\
     &  (arcsec)   & ($10^9L_\odot$) & (mag.)      & (mag)&(mag.)& ($10^9M_\odot$)& ($10^9M_\odot$) \\
\hline
Arp 194 S&11&16&0.38&0.83&0.4&45.6&18.9\\
Arp 194 Na (inner region)&5&4.4&0.56&1.19&0.1&-&10.0\\
Arp194 Nb (inner region)&4&3.8&0.78&1.40&0.15&-&18.8\\
A&3&0.97&0.05&0.35&0.225&$\leq$0.20&0.37\\
B&5&1.8&-0.03&0.42&0.15&$\leq$4.50&0.50\\
C&4&12&0.48&1.06&0.15&-&21.0\\
\hline\hline
\end{tabular}
\end{center}
\end{table*}

\subsection{Emission regions between the galaxies}
Emission-bright regions A and B represent the brightest islands of current star-formation between the galaxies. The ionized gas velocity dispersion of the region A does not exceed 30 $\textrm{km~s$^{-1}$}$ (see Fig. \ref{fig3}), which is significantly lower than for most of the  other parts crossed by slits  and is quite typical for star-forming galaxies where the velocity dispersion reflects the inner motions of gas in H{\sc ii} regions \citep{Moiseevetal2015}. Island A is a compact formation consisting  of several very close almost merging bright knots which are barely resolved even for the HST angular resolution.  It has a contrast shape elongated toward the centre of the S-galaxy. Region B looks quite different, it is more diffuse and contains separate knots evidently related to the insular nests of young stars. Internal  dynamics of these two regions is also different. Island A looks rather dynamically detached (a short plateau in the profiles with PA=115\degr, PA=138\degr, see Figs \ref{fig2} and \ref{fig3}). Its systemic velocity  slightly changes within 40 km/s along  both slits, crossing it. The dynamical mass of this region appears to be even lower than the stellar mass estimate (note however that the stellar model gives the accuracy of the mass estimate within  a factor of about 1.5 due to uncertainty of stellar initial mass function, see e.g. \citealt{McGaughSchombert2014}), thus leaving no room for dark matter, which is quite expectable for tidal dwarf. 

On the contrary, in the region B velocity dispersion of gas jumps from point-to-point, reaching 80 $\textrm{km~s$^{-1}$}$ and even higher values. Its line-of-sight velocity changes by more than 100 $\textrm{km~s$^{-1}$}$ smoothly without any steps while passing to the nearby areas.  It can indicate that the region B  is not gravitationally bound. Its virial mass, formally calculated, (see Table \ref{tab_phot}) confirms  this conclusion, if not to assume that it is a dark matter dominated dwarf.  

The colour of both regions - A and B - reveals a very young stellar population formed in the gaseous bridge between the merging galaxies. Their luminosities are also comparable (see Table \ref{tab_phot}). The oxygen abundance in these emission islands (12+$\log(\mathrm{O/H})\approx 8.25-8.35$ for O3N2 method) is rather high, being slightly lower (by $0.1-0.3$ dex) than in the parent galaxies, however it is typical for peripheral regions of spiral galaxies. It definitely indicates that the gas was chemically  recycled  before it was lost by galaxies. 

\begin{figure}                                                               
 \centering                                                                   
 \includegraphics[angle=-90,width=\linewidth]{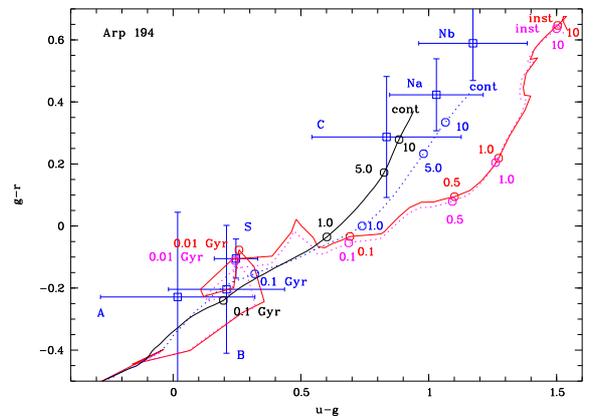}       
  \caption{ The $g-r$ versus $u-g$ diagram. The straight and dotted lines show the models for Salpeter and Kroupa IMFs correspondingly. We considered models for continuous star-formation (labeled as cont) and the models of instantaneous burst of star-formation (labeled as inst). The symbols with error-bars denote the regions marked  as in Fig.~\ref{fig1}.}
\label{ugr} 
\end{figure}

The absence of noticeable signs of old stars in the observed emission  regions between the galaxies was claimed earlier in \cite{Marziani2003}. To clarify a possible age of the regions we compared their position  on the $(g-r)-(u-g)$  diagram with evolutionary tracks computed with PEGASE2 software package\footnote{The diagram was kindly prepared by E. Egorova} in Fig.\ref{ugr}. Colour indexes were determined using the photometric data from SDSS survey and corrected for internal extinction. We considered tracks for $Z=0.004$, that is for metallicity several times lower than the solar value. Two solid lines with denotations of age represent the models for continuous star-formation (the shorter line, labeled as cont) and for instantaneous star-formation (the longer line, labeled as inst) for Salpeter IMF (solid lines) and for Kroupa IMF (dotted lines).  The position of two components of N- galaxy confirms their great age. In the case of S-galaxy the spectrum is defined mostly by young stellar population with the age lower than $10^8$ yr. However, it reflects the current intense burst of star-formation rather than the overall youth of the galaxy. The high luminosity of this galaxy in K-band ($L_K \approx4\cdot10^{10}L_\odot$) as well as relatively symmetric shape of the outer isophotes evidences in favour of long history of the galaxy preceded the observed star-formation burst.

Colour indexes of A and B- regions confirm the conclusion of their  recent formation. Unfortunately in the area of small ages a colour-colour diagram does not allow to estimate accurately the age and distinguish between the instantaneous and continuous star-formation. We can only argue that the age of stellar population in these regions is between $10^7$ and  $10^8$ yr. We also compared our data with Starburst99 model tracks (not shown here) and came to the same conclusions as for PEGASE2.

It is worth noting that, although both S and N- galaxies have very close line-of-sight velocities,  the velocity of gas  between galaxies along any slit we used is negative with respect to these galaxies, that is directed toward the observer. It evidences that the gas moves out of the sky plane, because its line-of-sight velocity increases with the distance to the nearest galaxy. One may conclude that the  farther   gas is  from S- or N-galaxy, the  higher it  raises above this plane.  Most likely, the emission islands between galaxies form a kind of a giant  arc facing its convex side to the observer. A special case presents the area adjacent to  S-galaxy which includes the region A:  its observed velocity is a little bit higher than the velocity of the centre of S –galaxy.  It agrees with the scenario where this young stellar island and its surrounding gas  falls into the disc of S-galaxy (see below). 

The elongated shape of the island A results evidently from the tidal stretching of stellar system residing in the free fall trajectory in the gravitational field of the massive S-galaxy. This circumstance can be used also to put some constrains on its age in terms of the simple kinematical model.   

The observed axis ratio of the island A is $a/b \approx2$. Let us assume that its elongation  is due to the tidal acceleration $a_t$ along the major axis $R$, which caused the increase of initial size $d_0 $ by the value  $\Delta d$, comparable to $d_0$. Tidal acceleration of expansion of the island along the axis of symmetry: 
\begin{equation}
a_t \leq d(V ^2/R)/dR \cdot(d_o+ \Delta d),
\end{equation} 
 where $V $ is the circular velocity at the periphery of S- galaxy at the distance R to the A-region (the inequality sign reflects the neglecting the self-gravity of the object). For the time needed the object to be stretched at $\Delta d\approx d$, we have:
\begin{equation}
t_{exp} \leq (2 \Delta d/a_t)^{1/2} \sim (R_0 / \sin \alpha)/( \sqrt{2} V), 
\end{equation}
where $R_0\leq R$ is the distance between the island A and S-galaxy in the sky plane, $\alpha$ is the angle between the spatial distance R and the line-of-sight. For $V\approx 150-200$ $\textrm{km~s$^{-1}$}$ and  $ d_0\approx \Delta d$, $R_0\approx $7 kpc we get $t_{exp} \leq (2-3)\cdot 10^7$  yr. 

Monotonous variation of the observed velocity along the slits passing through the chain of emission regions  between galaxies, as well as the trend of decreasing of oxygen abundance O/H with the distance from N and S-galaxies agrees with the conclusion that the emission  regions in the bridge of Arp194 were formed in the flow(s) of gas lost by  both galaxies after collision. The borderline between gas presumably lost by S and N-galaxies may be tentatively accepted as passing near the region B. As far as the radial gradient of O/H is inherent for spiral galaxies, those regions which now observed nearer to their parent galaxy apparently were closer to its centre before the collision. A modulo of radial gradients of O/H of spiral galaxies  usually lies in a range $0.01-0.05\ \mathrm{dex\ kpc^{-1}}$ (see, e.g. \citealt{Kudritzkietal2015}). The difference of $\log(\mathrm{O/H})$ along the slit PA=115\degr  ~between S- galaxy and the region B is roughly 0.15 dex (see Fig.~\ref{abund115}). A similar drop of abundance is between region B and Nb-galaxy (slit PA=138\degr). It corresponds to the difference of radial distances of their initial locations in a parent galaxy  $\sim$ 3--15 kpc, which is comparable with the radii of galaxies. Hence there is no need to seek for  mechanism of dilution of the observed emission gas by non-abundant medium. 

Numerical simulations of collision of gas-rich galaxies show that the gaseous disc which has undergone the off-centre collision develops a large-scale density wave having a shape of expanding asymmetric ring or asymmetric arc (in the case of high impact-factor) \citep[see, e.g. ][]{Gerberetal1992, MapelliMayer2012, Struck1997}. \cite{Struck1997} also considered the formation of a bridge between the collided galaxies in addition to the arc/ring. The bridge in their numerical hydrodynamic simulations is not tidal but rather hydrodynamic.  Gas splashed out from  discs remains gravitationally bound with a parent galaxy and eventually falls back into it. Evidently the gas located between galaxies of Arp194 will also share this  fate being eventually accreted by the galaxies.

The line-of-sight velocity of island A is by $15-25\ \textrm{km~s$^{-1}$}$ higher than that of the centre of S-galaxy which proves its fall into the galaxy. Since the free fall speed onto the galaxy should be higher by an order of magnitude, one can conclude that the velocity vector of the island forms a small angle with the image plane (of course, its angle to the plane of the disc is significantly higher). A current dip of stellar island A into S-galaxy is well illustrated by the colour HST image, where  one can distinctly see against the background of  S- galaxy that this island is fragmenting into pieces and  is connected with the  reddish dust areas of the depressed surface brightness projected onto the disc. 

Unfortunately, there is no information on the distribution of neutral atomic gas in the bridge between the galaxies of Arp194. However, the existence of regions containing young stars definitely evidences in favour of the presence of a significant amount of H{\sc i} both in the galaxies and between them. A formation of the extended regions of young stars is the consequence of compression of gas thrown out of the galaxies. In general case, two processes can be involved in this process (which may act jointly): a gravitational instability in the case of low gas velocity dispersion  in the densest gaseous clumps of the bridge, and a  shock collision of gaseous streams in strongly perturbed medium which arose during the passage of a galaxy through another one. At any rate, the formation of gravitationally bound stellar islands (tidal dwarfs) requires a special conditions. 

As we have already noted in Introduction, the extended emission clumps are quite seldom observed between the close  interacting galaxies. They more frequently occur not in a bridge, but rather in tidal tails where the turbulent motions of gas have a time to fade away, and the velocity of gas forming a tail falls with distance from a galaxy which leads to an increase of its density. Active star formation in a bridge demands special conditions.

As we noted above, in the case of Arp 194 in the regions of low intensity of emission lines, where there are no bright extended star-forming regions the gas appears to be well mixed chemically since the variation of O/H is very low there (see Fig.~\ref{abund138}): the oxygen abundance $12+\log(\mathrm{O/H})$ remains within the range of $8.3-8.4$. The stirring of rarefied gas confirms the presence of large-scale motions of gas which are naturally to expect in a case of collision of gas medium of two interpenetrating galaxies. Shock waves support the turbulent motions of gas which could lead both to the high velocity dispersion of gas in local regions and to the shock excitation of atoms.  Turbulent velocities die out slower in the regions of rarefied gas. Indeed, the regions with higher velocity dispersion of gas ($ > 70$ $\textrm{km~s$^{-1}$}$) correspond to the areas along the slit where the line intensity is the lowest and there are no noticeable regions of star-formation (Figs. \ref{fig3}, \ref{abund138}). In particular, it may be applied to the wide zone to the east from Na+Nb galaxies, crossed by the slit with PA=138\degr~(see Fig. \ref{fig3}).

The presence of high turbulent velocities of gas in the bridge which energy is supplied by shock waves was also found for another  system -- VV254 (Taffy), consisted of two close galaxies that have undergone head-on collision \citep{Zhuetal2007}. The high velocity dispersion of gas in the bridge between the galaxies exceeding 100 $\textrm{km~s$^{-1}$}$ was measured by infra-red emission lines of warm molecular hydrogen by \cite{Petersonetal2012}.  It is also reproduced in the numerical simulations of the collision  \citep{Vollmeretal2012}. At the same time, there is almost no star-formation between the Taffy galaxies  despite of the large amount of cold gas (H{\sc i}+H$_2$) with a total mass of $~7\cdot10^9\mathrm{M}_\odot$ in the bridge, even not taking into account a warm H$_2$. The observed length and width of Taffy' bridge is about  $\sim10$~kpc. The extension of the bridge along the line-of-sight is unknown, but it unlikely exceeds the distance between the galaxies (12 kpc). Assuming that the volume filled by gas is equal to $10^3\ \mathrm{kpc}^3$ we get the mean volume density of directly observed gas $\sim 5\cdot10^{-25}\mathrm{g~cm^{-3}}$ which is comparable to that in the plane of our Galaxy. Evidently the gas density is not homogeneous, so that its local values can be significantly higher. Too high velocity dispersion of gas thrown out of the galaxies can  prevent the appearance of large star-formation clumps in this system.

The numerical simulations show that the collision of galaxies in Taffy took place $\sim 2\cdot 10^7$ yr ago \citep{Vollmeretal2012}. In comparison, Arp194 is observed at later stage of evolution ($\sim 10^8$ yr) which is comparable with the expected time of free fall of gas from the bridge to the galaxies.  The absence of regions of intense star-formation between closely interacting galaxies in such systems as Taffy is apparently a result of short duration of the stage between the beginning of active star-formation in the bridge after the damping of strong turbulent motions and the falling of gas from the bridge back to the parental galaxies. Regions, similar to A or B, observed in Arp 194, have not yet been formed there.

\section{Conclusions}\label{conclusion} 
We performed the long-slit spectral observations of interacting system Arp194 using three positions of the slit. The  main results of data processing and the analysis  are given below.
\begin{itemize} 
\item We obtained the distribution of the line-of-sight velocity and velocity dispersion as well
as the intensities of emission lines and oxygen abundance $12 + \log(\mathrm{O/H})$ along the slits.
\item A special attention was paid to bright condensations between galaxies: the extended star-forming island (region A), multi-component  site of star-forming knots (region B) and the compact object C (see Fig.1). We conclude that the region A is gravitationally bound short-lived tidal dwarf  galaxy which is falling into the massive southern galaxy. The comparison of our dynamical estimate of mass with that followed from the photometry indicates that it is devoid of a considerable amount of dark matter. Region B is apparently not in the gravitational equilibrium.  Object C appeared to be a background spiral galaxy.   
\item  The gas in the system is only partially chemically mixed: the regions with low intensity of emission lines, crossed by the slits, do not reveal a significant systemic variation of O/H.  At the same time, we observe the tendency of O/H to decrease  with galactocentric distances both from S and N galaxies. Local velocity dispersion  exceeding 50 $\textrm{km~s$^{-1}$}$ mostly takes place in the diffuse gas devoid of the extended regions of star-formation evidencing a strong turbulent motions of gas. 
The velocity and the abundance distributions allow to conclude that the gas was stripped from both galaxies and avoided a strong mixing. Most likely it will return  back to their parent galaxies rather than it follows a cross-fueling scenario advanced by  \cite{Marziani2003}.  
\item We compared the colour indexes of several discrete regions of star-formation (including the central parts of the main galaxies of Arp194) with evolutionary tracks using both PEGASE2 and Starburst99 models. We  found that the colour of northern galaxies corresponds to the old stellar systems; the colour of the southern galaxy gives evidence that it experienced a strong  burst of star-formation; regions A and B  have the age of stellar population of $10^7-10^8$ yr.
\item   System Arp 194 is not typical, because it has the extended fireplaces of intense star-formation between the galaxies. We propose that the rare occurrence of regions of star-formation in the bridges between closely interacting galaxies is the result of short duration of the stage between the fading of strong turbulent motions of gas caused  by strong interaction and the accretion of gas from the bridge back to the parent galaxies.

\end{itemize}

\section*{Acknowledgements} 
We thank the anonymous referee for the valuable comments.
 Authors are grateful to E. Egorova for comparison of the colour indexes of the regions with model evolutionary tracks on colour-colour diagram. The observations at the 6-meter BTA telescope were carried
out with the financial support of the Ministry of Education
and Science of the Russian Federation (agreement No. 14.619.21.0004, project ID RFMEFI61914X0004). In this study, we used the SDSS DR9 data. Funding
for the SDSS and SDSS-II has been provided by the Alfred P. Sloan Foundation,
the Participating Institutions, the National Science Foundation, the U.S.
Department of Energy, the National Aeronautics and Space Administration, the
Japanese Monbukagakusho, the Max Planck Society, and the Higher Education
Funding Council for England. The SDSS Web site is http://www.sdss.org/.
Based on observations made with the NASA/ESA Hubble Space Telescope, obtained from the data archive at the Space Telescope Science Institute. STScI is operated by the Association of Universities for Research in Astronomy, Inc. under NASA contract NAS 5-26555.

This work was supported by Russian Foundation for Basic Research (RFBR) 14-22-03006-ofi-m. The Russian Science Foundation (RSCF) grant No. 14-22-00041 supported the study of gas chemical abundance and kinematics. 
AS and OE are thankful to the RFBR grant No. 15-32-21062; AS is grateful to the RFBR grant No. 15-52-15050 and the Russian President's grant No. MD-7355.2015.2 for support. 
\bibliographystyle{mnras}
\bibliography{Arp194}

\label{lastpage}

\end{document}